\documentclass[letterpaper]{mc2023}
\usepackage{float}
\usepackage[caption=false]{subfig}
\usepackage{tabls}
\usepackage{cites}
\usepackage{epsf}
\usepackage{appendix}
\usepackage{ragged2e}
\usepackage[top=1in, bottom=1in, left=1in, right=1in]{geometry}
\usepackage{enumitem}
\usepackage{xcolor}

\setlist[itemize]{leftmargin=*}
\usepackage{caption}
\title{High-fidelity treatment for object movement in\\time-dependent Monte Carlo transport simulations}
\author{%
  \textbf{Ilham Variansyah$^{1}$ and Ryan G. McClarren$^{2}$}\vspace{3pt} \\
  \vspace{6pt}\\
  $^1$School of Nuclear Science and Engineering\\
  Oregon State University,
  Merryfield Hall, Corvallis, OR 97331 
  \vspace{6pt}\\
  $^2$Department of Aerospace and Mechanical Engineering\\
  University of Notre Dame,
  Fitzpatrick Hall, Notre Dame, IN 46556 \vspace{6pt}\\ 
  {variansi@oregonstate.edu}
}
\newcommand{\authorHead}{Ilham Variansyah and Ryan G. McClarren}
\newcommand{\shortTitle}{High-fidelity object movement in transient MC}
\begin{document}
\maketitle
\justify 
\parskip 6pt plus 1 pt minus 1 pt

\begin{abstract}
We investigate the use of time-dependent surfaces in Monte Carlo transport simulation to accurately model prescribed, continuous object movements. 
The performance of the continuous time-dependent surface technique, relative to the typical stepping approximations and the recently proposed at-source geometry adjustment technique, is assessed by running a simple test problem involving continuous movements of an absorbing object. 
A figure of merit analysis, measured from the method's accuracy and total runtime, shows that the time-dependent surface is more efficient than the stepping approximations. 
We also demonstrate that the time-dependent surface technique offers robustness, as it produces accurate solutions even in problems where the at-source geometry technique fails. 
Finally, we verify the time-dependent surface technique against one of the multigroup 3D C5G7-TD benchmark problems.
\end{abstract}
\vspace{6pt}

\section{INTRODUCTION}

Many time-dependent transport problems involve prescribed continuous object movements.
Some practical examples include criticality experiments (e.g., Honeycomb and Lady Godiva~\cite{criticality_accident}), reactor control element movements, and transient analyses of innovative reactor concepts, such as the coupling/decoupling of the Holos-Quad core's Subcritical Power Modules~\cite{stauff2019}.

Time-dependent transport problem is typically modeled into a series of steady-state problems, where continuous object movements are approximated into steps, introducing discretization error.

This stepping approximation of object movement has been traditionally adapted not only in deterministic transport methods~\cite{shen2019,ryu2017} but also in time-dependent Monte Carlo (MC)~\cite{shaukat2017,faucher2018,molnar2019}.
A sufficiently small time step is required to minimize the discretization error. Such a constraint is inherent in deterministic methods---but in MC, it comes with an additional consequence.
Smaller time steps mean a more frequent particle census, which is required to pause the running MC simulation and subsequently change the model; this particle synchronization, however, adds to the overall simulation run time. 
There are several good reasons for particle census in time-dependent MC simulation, including (1) integrating non-linear feedback, (2) outputting large time-dependent solution tallies, (3) controlling population size, and (4) re-balancing the workload of parallel processors. 
Determining optimal census frequency or time step size is necessary to achieve efficient time-dependent MC simulation; implementing the stepping approximation to model continuous object movement adds a constraint in determining such optimal value.
Regardless, one of the core values of MC simulation is its capability of high-fidelity modeling---it is rather unfortunate if we have to introduce some discretization errors on something that MC can resolve continuously.

An alternative technique for modeling object movement was recently proposed~\cite{durkee2016}.
The main idea (and assumption) is that the entire model geometry is adjusted at the creation of source and delayed particles.
This technique, hereinafter referred to as ``at-source geometry adjustment'', is reasonably effective for many applications. 
However, its accuracy would deteriorate as the particles' velocities become comparable with the objects' and the spacings between the particles and the objects increase~\cite{durkee2016}.

In this paper, we investigate the use of time-dependent surfaces to achieve high-fidelity modeling of prescribed, continuous object movements in MC simulations. 
In Section~\ref{sec:td_surf}, we formulate and discuss the implementation of the time-dependent surfaces. 
Then, in Section~\ref{sec:simple}, we devise a simple test problem to asses the relative performances of the proposed technique and the existing ones---i.e., stepping and at-source adjustment~\cite{durkee2016}. 
In Section~\ref{sec:c5g7}, we demonstrate the use of the time-dependent surfaces in solving one of the C5G7-TD benchmark problems~\cite{hou2017}. 
Finally, Section~\ref{sec:summary} summarizes and discusses future work.

\section{TIME-DEPENDENT SURFACE}\label{sec:td_surf}

In the MC method, the domain of the transport problem is typically modeled into material cell objects bounded by constructive solid geometry (CSG) surfaces. 
Changing the position or angle of the bounding surfaces effectively moves or changes the geometry of the associated objects. 
CSG surface is usually defined by the constants of the quadric surface equation:
\begin{equation}
\mathcal{S}(x,y,z)=Ax^2+By^2+Cz^2+Dxy+Exz+Fyz+Gx+Hy+Iz+J,
\end{equation}
and if we make the relevant constants time-dependent, we effectively move or change the surface. 

The axis-aligned planes are the simplest yet widely useful case of a time-dependent surface. 
As an example, the time-dependent CSG equation for the plane surface parallel to the $z$-axis is
\begin{equation}
\mathcal{S}(z,t)=z+J(t).
\end{equation}
To determine on which side a particle is with respect to the moving surface (which is needed to determine particle cell), we can evaluate the sign of $\mathcal{S}(z_0,t_0)$, where $z_0$ and $t_0$ are respectively the particle position and time. This is the only modification needed should one uses a delta tracking algorithm~\cite{delta_tracking}. However, in surface tracking, we also need to determine particle flight distance to the surface $l_\mathrm{surf}$ by solving
\begin{equation}
\mathcal{S}(z_0+l_\mathrm{surf}u_z,t_0+l_\mathrm{surf}/v)=0,
\end{equation}
where $u_z$ and $v$ are respectively the $z$-component direction and the speed of the particle. 
For a constant speed surface, $J(t)=J_1t+J_0$, this would be
\begin{equation}
l_\mathrm{surf}=-\frac{\mathcal{S}(z_0,t_0)}{u_z+J_1/v}.
\label{eq:l_surf}
\end{equation}
We note that if the surface speed $J_1=0$, Eq.~\eqref{eq:l_surf} reduces to the static surface formula.

It is worth mentioning that this simple time-dependent axis-aligned plane surface may be enough to model many cases of moving objects in transport problems, from typical reactor control rod insertion/withdrawal to the coupling/decoupling of the Subcritical Power Modules of the Holos-Quad concept~\cite{stauff2019}---i.e., by bounding the module (or assembly) universes within the time-dependent surfaces.

\section{SIMPLE TEST PROBLEM}\label{sec:simple}

\begin{figure}[!htb]
  \centering
  \includegraphics[scale=0.25]{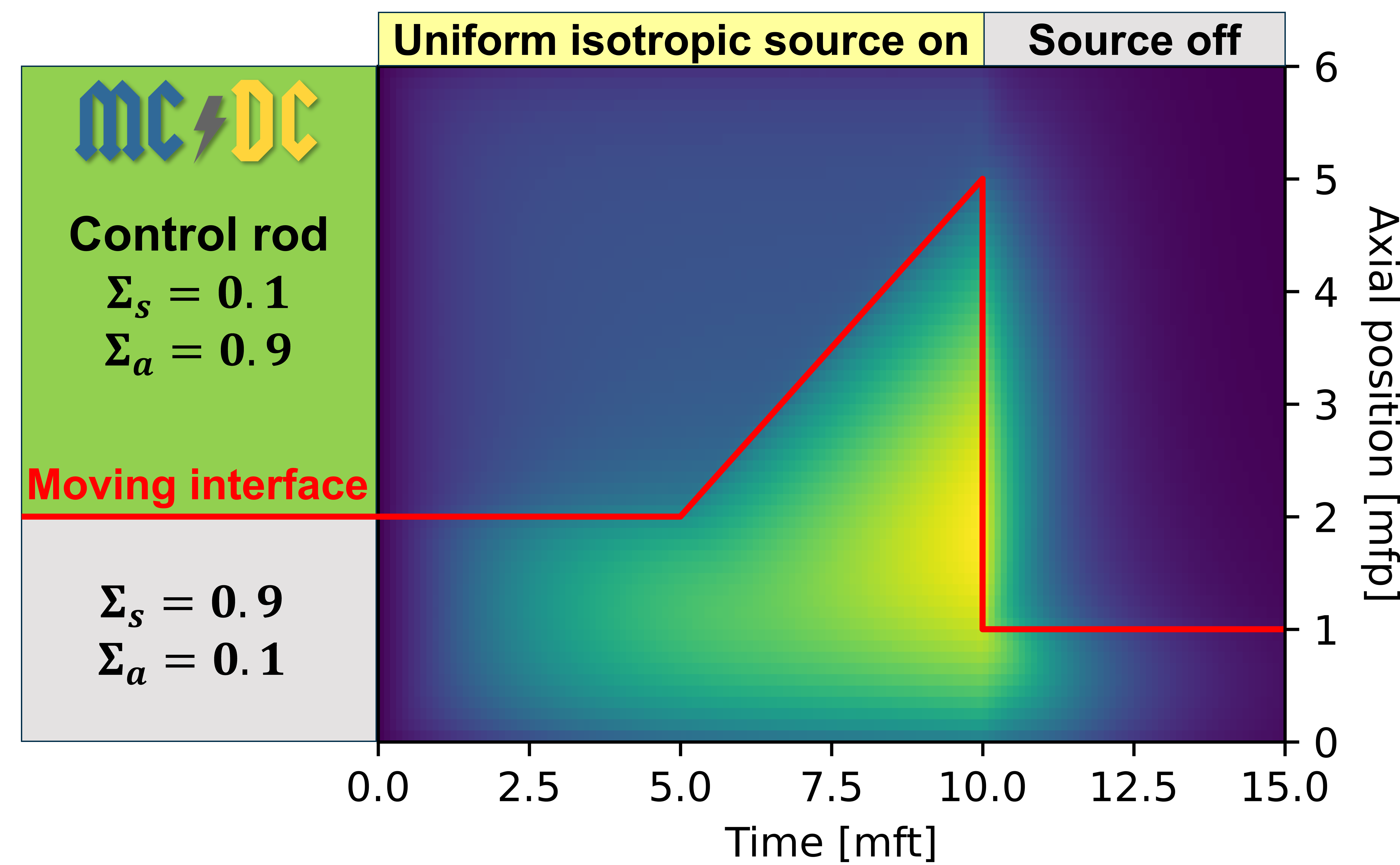}
  \caption{The test problem scalar flux reference result ($N_\mathrm{hist}=10^{10}$).}
  \label{fig:test_problem}
\end{figure}

To test and assess the benefits of using the proposed time-dependent surfaces, let us consider a simple mono-energetic two-region slab problem, where we continuously change the position of the interface separating the absorbing ($\Sigma_a=0.9$, $\Sigma_s=0.1$) and scattering ($\Sigma_a=0.1$, $\Sigma_s=0.9$) materials. 
The red line in Fig.~\ref{fig:test_problem} highlights the interface position as a function of time. 
The domain spans $z\in[0,6]$ with vacuum boundary conditions, the simulation starts with a zero initial condition, a uniform fixed source isotropically emits particles in $t\in[0,10]$, and the simulation ends at $t=15$. 
All constants and variables are in the unit of the mean free path (mfp) and the mean free time (mft). We note that the problem tries to demonstrate typical control rod withdrawal and insertion.

\begin{figure}[!htb]
  \centering
  \includegraphics[scale=0.25]{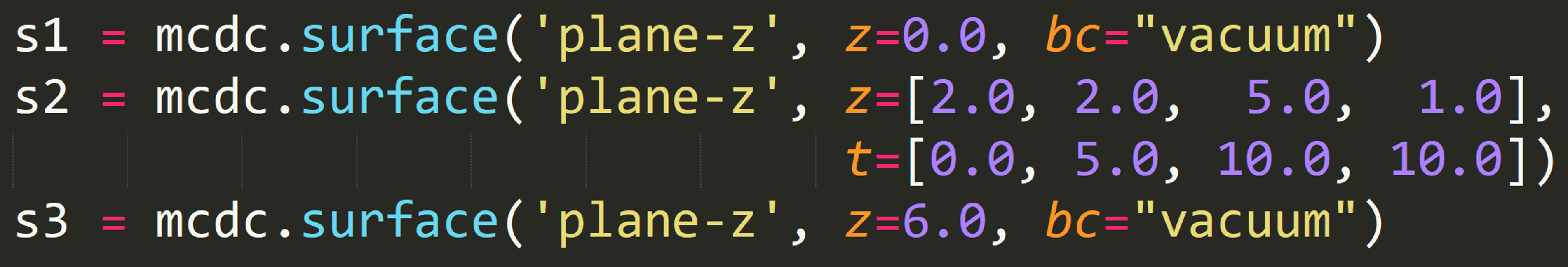}
  \caption{A user interface example for setting the time-dependent surface of the test problem.}
  \label{fig:input}
\end{figure}

We implement the time-dependent axis-aligned plane surface formulated in Section \ref{sec:td_surf} (as well as the stepping approach and the at-source geometry adjustment technique~\cite{durkee2016}, for performance comparison) into the Python-based MC code MC/DC\footnote{https://github.com/CEMeNT-PSAAP/MCDC}~\cite{variansyah2023}. 
The time-dependent surface position is modeled as a piece-wise linear function and (as an example of a possible user interface) set up in the input file as shown in Fig.~\ref{fig:input}. 
The test problem is run in analog using the time-dependent surface capability with $10^{10}$ histories, and the scalar flux result---calculated via track-length estimator in a uniform mesh grid $dz=dt=0.1$---is shown in Fig.~\ref{fig:test_problem}. 
We use this highly-precise solution as the reference for the performance comparison in the rest of this section.

Besides the highly-precise calculation of the reference solution, we solve the test problem with a significantly lower, but still reasonably high, number of histories of $10^8$ on 360 processors on LLNL’s compute platform Quartz (Intel Xeon E5-2695).
Four simulation cases---different on how the moving interface is modeled---are considered: (1) using the continuous time-dependent surface capability, (2) using the stepping approximation, (3) using the stepping approximation with material mixing, also known as ``decusping", and (4) using the at-source geometry adjustment technique~\cite{durkee2016}.

In Case 2 and Case 3, we consider 1, 2, 4, 8, 16, and 32 steps during the interface ramp change ($t\in[5,10]$), yielding MC simulations with $n+1$ censuses, where $n$ is the number of steps. 
The census is performed only to change the interface position (no population control or other operation is performed). 
The stepping is made in implicit-Euler style, where 1 step means instantaneously moving the interface from $z=2$ to $z=5$ at $t=5$. In Case 3, however, the homogenized mixture of the two materials ($\Sigma_a=0.5$, $\Sigma_s=0.5$) is used to model the continuous transition in each step. 
It is worth mentioning that in the implementation, some of the fixed-source particles may be emitted beyond a currently active census time. 
In that case, the particles are stored straight into the census bank and will start being transported only when their respective time tags are within the active census time.

\begin{figure}[!htb]
    \centering
    \subfloat[Stepping, 2 steps]{%
    \resizebox*{6cm}{!}{\includegraphics{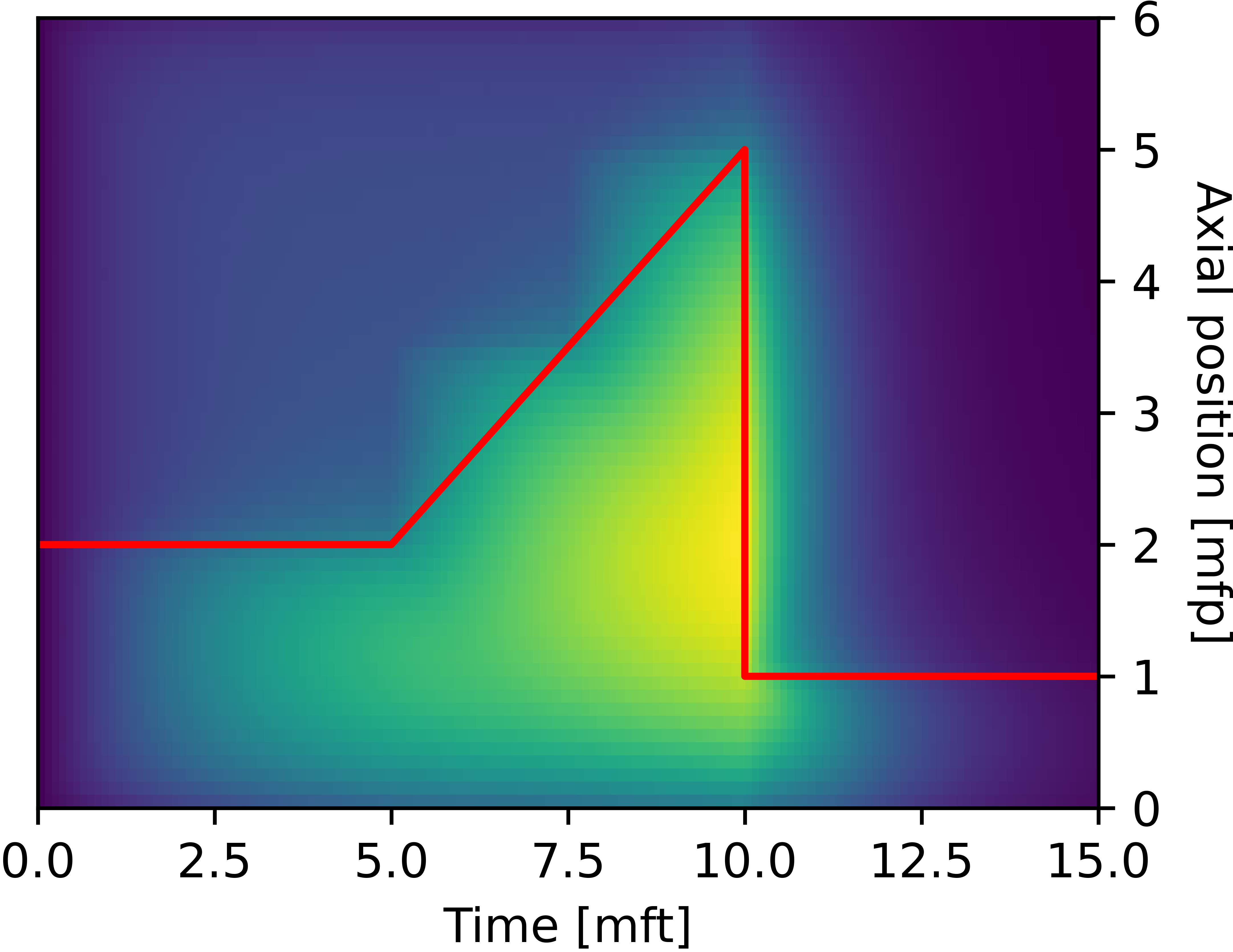}}}\hspace{5pt}
    \subfloat[Stepping with mixing, 2 steps]{%
    \resizebox*{6cm}{!}{\includegraphics{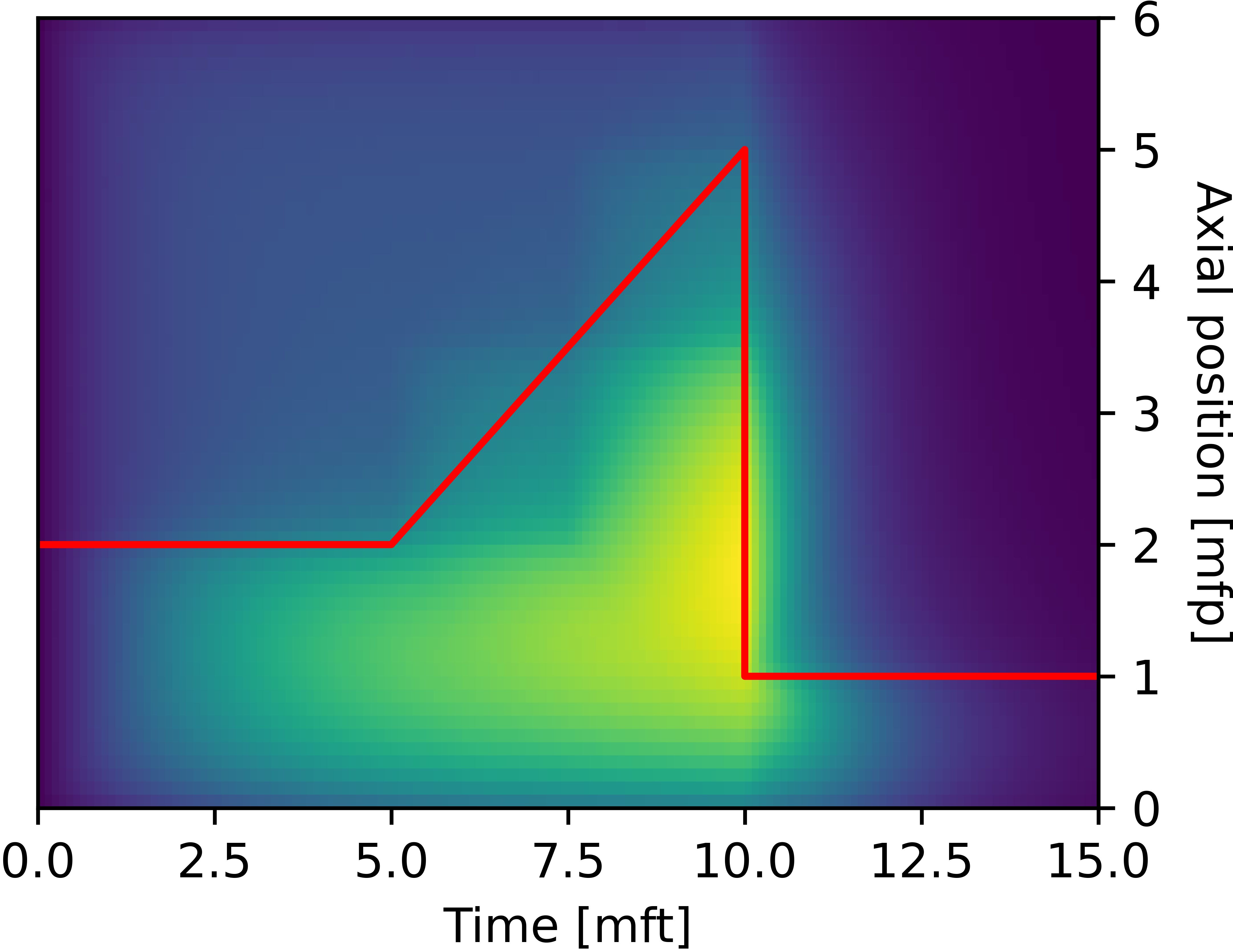}}}\\
    \subfloat[At-source geometry adjustment]{%
    \resizebox*{6cm}{!}{\includegraphics{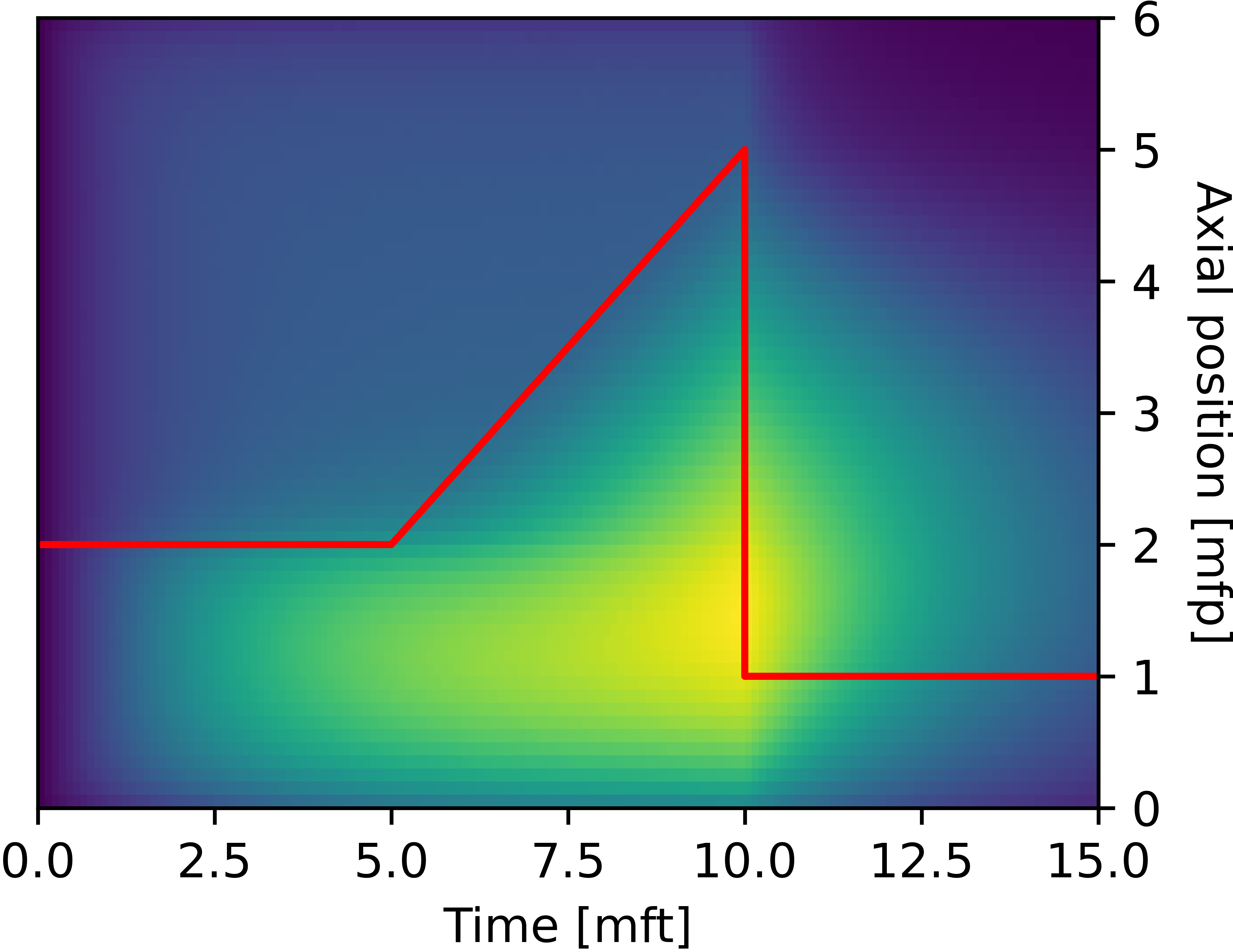}}}
    \caption{Test problem scalar fluxes obtained with the stepping and at-source adjustment techniques ($N_\mathrm{hist}=10^{8}$).} \label{fig:coarse_result}
\end{figure}

Figure~\ref{fig:coarse_result} shows the results of Cases 2, 3, and 4. 
The result of the time-dependent surface technique Case 1 is not shown as it is very similar to the reference solution in Fig.~\ref{fig:test_problem}, except for the minor statistical noise. 
Subfigures~\ref{fig:coarse_result}(a) and \ref{fig:coarse_result}(b) demonstrate how the stepping techniques discretize the continuous movement of the absorber, which introduce significant error if the number of steps is not sufficiently fine. On the other hand, Subfigure~\ref{fig:coarse_result}(c) shows that the at-source geometry adjustment technique gives a highly inaccurate result. 
This is because the devised test problem challenges the stipulations of the technique~\cite{durkee2016}. First, the absorbing material withdrawal speed is comparable to the particle speed. 
Second, some source particles are emitted very closely (in space and time) to the moving absorber. 
Third, half of the source particles are born before the absorbing material starts to move; particles born in $t<5$ will not feel the absorber withdrawal in $t\in[5,10]$, which causes the significant underestimation in $t\in[5,10]$. 
Finally, all of the source particles are born before the final absorber insertion at $t=10$; this causes the significant overestimation in $t\in[10,15]$. 

\begin{figure}[!htb]
    \centering
    \subfloat[2-norm of relative error]{%
    \resizebox*{6cm}{!}{\includegraphics{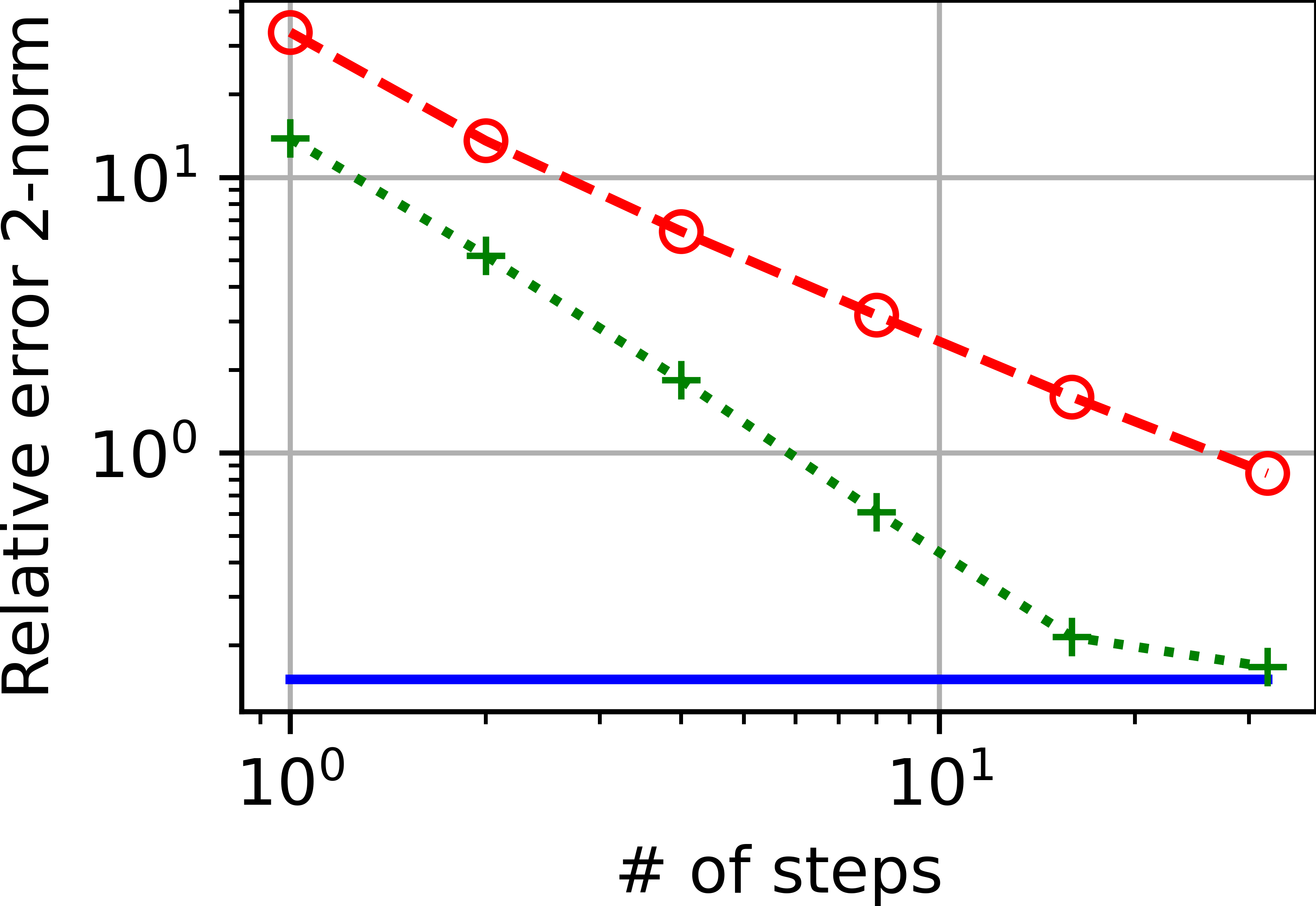}}}\hspace{5pt}
    \subfloat[Total runtime]{%
    \resizebox*{6cm}{!}{\includegraphics{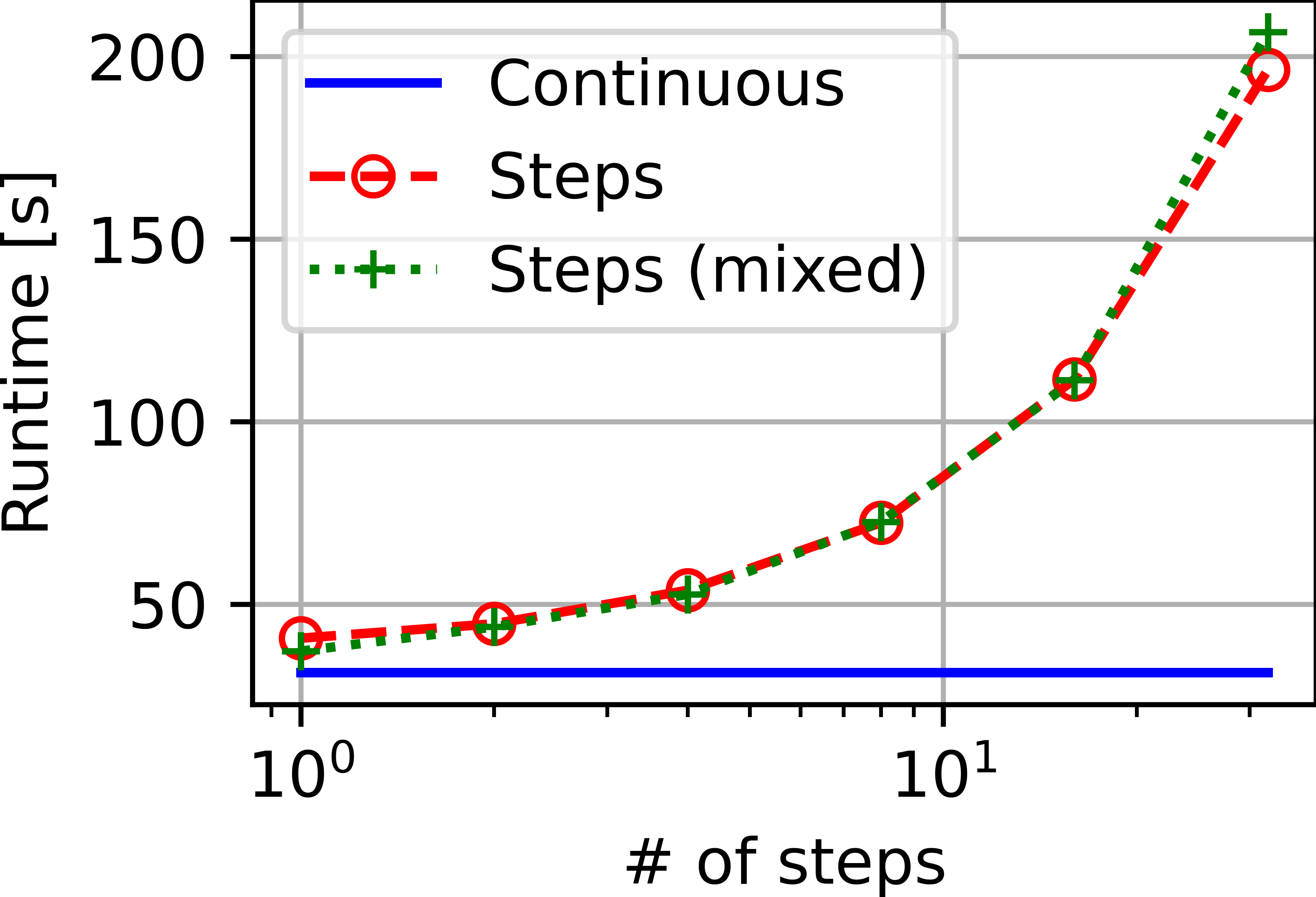}}}\\
    \subfloat[Figure of merit]{%
    \resizebox*{6cm}{!}{\includegraphics{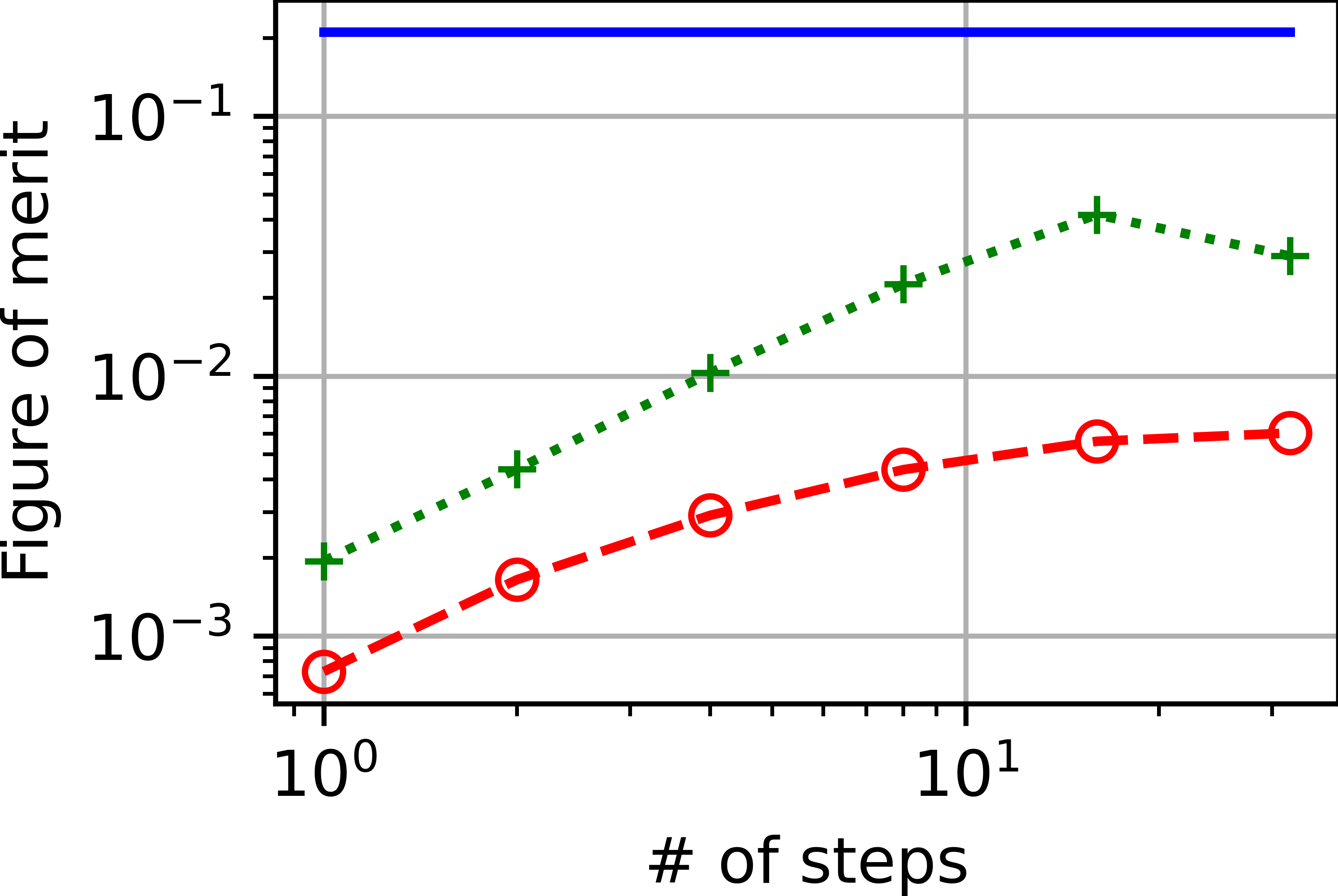}}}
    \caption{Performance metrics of the continuous time-dependent surface  (Continuous) and stepping techniques (Steps).} \label{fig:result}
\end{figure}

Next, we quantify the relative efficiencies of the continuous time-dependent surface technique and the stepping approximations. 
We consider two performance metrics: (1) 2-norm of scalar flux relative error in $t\in[5,15]$ and (2) simulation runtime. 
We also consider a figure of merit (FOM), defined as the inverse of the product of the two metrics. 
The resulting performance metrics are compared in Fig.~\ref{fig:result}. 
Subfigure~\ref{fig:result}(a) shows that the errors of the stepping approximations (Steps) reduce and approach the value of the continuous time-dependent surface technique (Continuous) as we increase the number of steps, where the error of the stepping approximation with material mixing reduces more rapidly before it eventually hits the error threshold. 
However, as we increase the number of steps, more censuses are performed, increasing the total run time of the stepping approximations, as shown in the Subfigure~\ref{fig:result}(b). 
Finally, it is observed from Subfigure~\ref{fig:result}(c) that the FOMs of the stepping approximations are always lower than the continuous time-dependent surface technique and would get worse as we further increase the number of steps because the error eventually stops improving while the run time keeps increasing.
We note that the runtime of the at-source geometry adjustment technique is similar to the time-dependent surface technique; however, its relative error 2-norm is much larger (over 2000 times).

\section{VERIFICATION AGAINST C5G7-TD BENCHMARK}\label{sec:c5g7}

In this section, we test our implementation of the time-dependent surface technique against one of the multigroup 3D C5G7-time-dependent benchmark problems~\cite{hou2017}. 
We consider exercise TD4-2 which involves continuous insertion and withdrawal of a control rod bank. 
With our implementation in MC/DC, we can model the tips of all the moving control rods with a single, shared time-dependent axis-aligned surface. Note that we also need to apply the same time-dependent surface to bind the water moderator below the control rods.

To run a MC simulation of this type of reactor transient (i.e., starting from a critical steady state), we need to prepare the initial-condition neutrons and delayed neutron precursors. 
We use the initial particle sampling technique proposed in~\cite{variansyah2023ic}. 
First, we run an accurate criticality calculation on the un-rodded configuration: with 50 inactive and 150 active cycles and 20 million particles per cycle, we get a $k$-eigenvalue of $1.165366 \pm 2.8$ pcm. 
We then follow the techniques detailed in~\cite{variansyah2023ic} to prepare the initial-condition particles. We set the neutron and delayed neutron precursor target sizes to be $5 \times 10^6$.

\begin{figure}[!htb]
  \centering
  \includegraphics[scale=0.8]{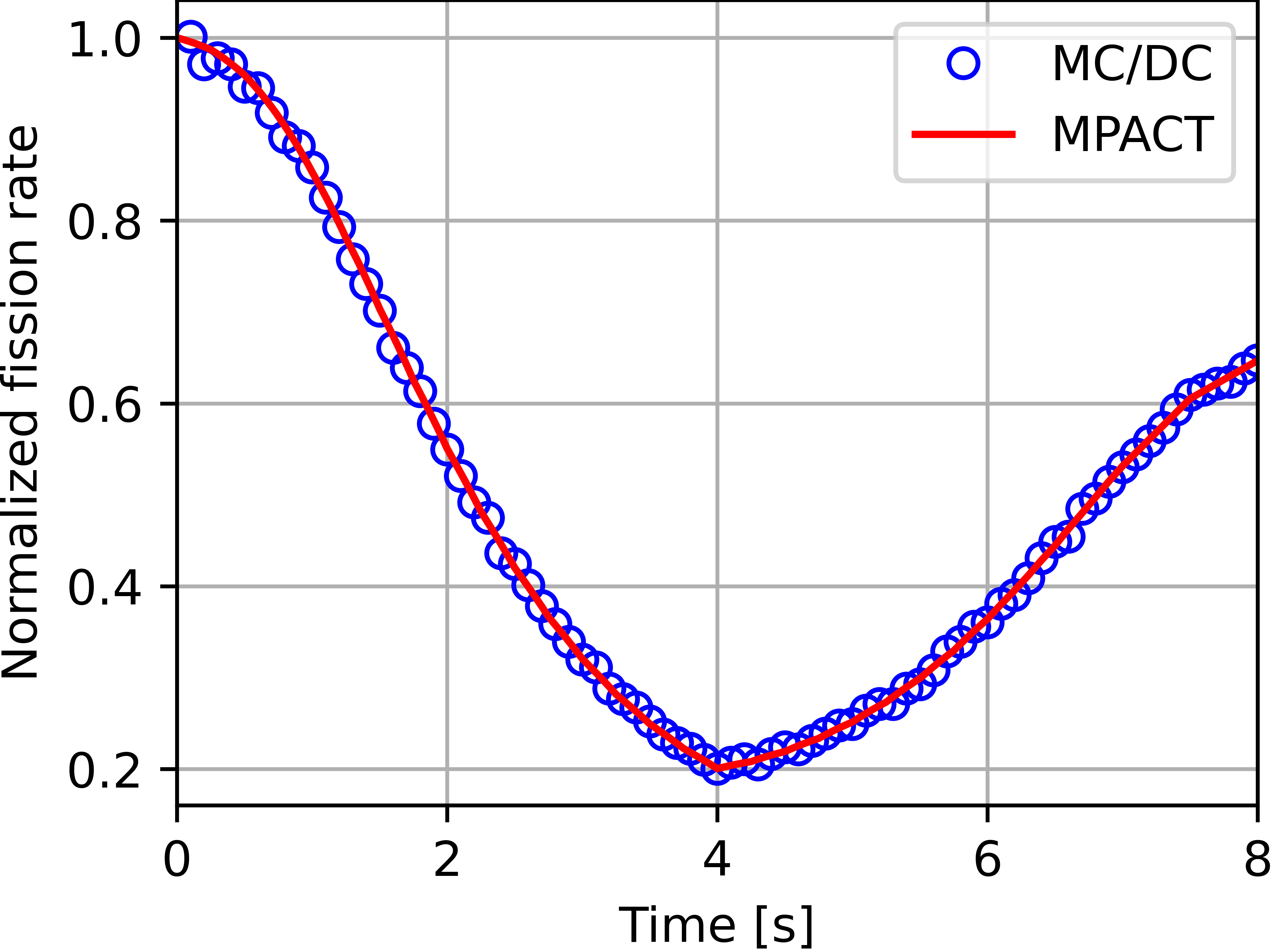}
  \caption{MC/DC result for the 3D C5G7 TD4-2 benchmark problem, run with the time-dependent surface technique with about 5 million initial neutrons and delayed neutron precursors and time-crossing tally estimator. MPACT result~\cite{shen2019} is also presented as a reference.}
  \label{fig:c5g7-td}
\end{figure}

With the sampled initial condition particles, we run the time-dependent MC problem. 
We record the total fission rate via a time-crossing estimator~\cite{variansyah2022pct} in a uniform time grid of $\Delta t=0.1$ s. 
We note that the simulation is run in ``analog"~\cite{variansyah2023ic}---i.e., with unit weight neutrons and without any variance reduction technique or population control. The result is shown in Figure~\ref{fig:c5g7-td}. It is evident that MC/DC result is in good agreement with the result generated by Shen et al.~\cite{shen2019} using the deterministic code MPACT with a uniform time step size of $\Delta t=0.025$ s, which is 4 times smaller than what is used in MC/DC. Different from deterministic codes, which need to resolve the continuously moving rods with sufficiently small $\Delta t$, MC simulation with time-dependent surfaces can use arbitrarily large $\Delta t$ without losing accuracy in modeling the transient objects.

\section{SUMMARY AND FUTURE WORK}\label{sec:summary}
We investigate the use of time-dependent surfaces for high-fidelity treatment of continuous object movement in MC simulations. This object-moving technique is an alternative to the widely-used stepping approximations and the recently proposed at-source geometry adjustment technique~\cite{durkee2016}.

We formulate the application of time-dependent axis-aligned plane surfaces for the surface-tracking algorithm, implement it into the MC code MC/DC~\cite{variansyah2023}, and verify it against a simple test problem and a 3D C5G7-TD benchmark problem. Through a figure of merit analysis of the test problem, we find that the time-dependent surface is largely more efficient than the stepping approximations. We also demonstrate that the time-dependent surface technique offers robustness, as it produces accurate solutions even in problems where the at-source geometry technique fails. 

Future work includes implementing more complex geometry changes, such as those involving translation, rotation, and expansion of quadric surfaces~\cite{durkee2016}. Also, mentioned a couple of times in the paper, it would be interesting to test the formulated time-dependent axis-aligned surface to model coupling and decoupling of the Holos Quad core's Subcritical Power Modules~\cite{stauff2019}.

Furthermore, the time-dependent surface technique can be synergized with time-dependent cross-section techniques---e.g., by adapting the sampling technique proposed by Brown and Martin~\cite{brown}---to model system expansion and compression. Some time-dependent transport problems are nonlinear (e.g., multi-physics simulations), which inevitably require performing particle time census to update the system's configuration. In this case, the use of the time-dependent surface technique, as well as its synergy with a time-dependent cross-section technique, can still be beneficial---as in that case, the techniques serve as a higher-order alternative to the typically implemented stepping (both in geometry and material property changes) approximations.

\section*{ACKNOWLEDGEMENTS}
This work was supported by the Center for Exascale Monte-Carlo Neutron Transport (CEMeNT) a PSAAP-III project funded by the Department of Energy, grant number DE-NA003967.

\setlength{\baselineskip}{12pt}
\bibliographystyle{mc2023}
\bibliography{mc2023}
\end{document}